# Computer-aided Diagnosis of Malaria through Transfer Learning using the ResNet50 Backbone

**SANYA SINHA[1*] AND NILAY GUPTA[2]**

*Dept. of Electronics and Communications Engineering, Birla Institute of Technology Mesra, Patna Campus, Patna, India*

*Centre for Artificial Intelligence, ZHAW School of Engineering, Winterthur, Switzerland*

*[*ssanya0904@gmail.com](mailto:ssanya0904@gmail.com)*

**Abstract**

According to the World Malaria Report of 2022, 247 million cases of malaria and 619,000 related deaths were reported in 2021. This highlights the predominance of the disease, especially in the tropical and sub-tropical regions of Africa, parts of South-east Asia, Central and Southern America. Malaria is caused due to the *Plasmodium* parasite which is circulated through the bites of the female Anopheles mosquito. Hence, the detection of the parasite in human blood smears could confirm malarial infestation. Since the manual identification of *Plasmodium* is a lengthy and time-consuming task subject to variability in accuracy, we propose an automated, computer-aided diagnostic method to classify malarial thin smear blood cell images as parasitized and uninfected by using the ResNet50 Deep Neural Network. In this paper, we have used the pre-trained ResNet50 model on the open-access dataset provided by the National Library of Medicine's Lister Hill National Center for Biomedical Communication for 150 epochs. The results obtained showed accuracy, precision, and recall values of 98.75%, 99.3% and 99.5% on the ResNet50 (proposed) model. We have compared these metrics with similar models such as VGG16, Watershed Segmentation and Random Forest, which showed better performance than traditional techniques as well.

**Key words:** Artificial Intelligence, Automated Diagnosis, Computer Vision, Deep Learning, Healthcare,

**Introduction**

Malaria is a fatal, mosquito-borne disease caused mainly due to the presence of the *Plasmodium falcifarum* parasite [1]. Malaria is typically symptomized with high body temperatures, chills, and body ache like most other influenza-adjacent diseases [2]. However, if left untreated, malaria may also cause severe health complications including seizures, neuro-disability, multiple organ failure, and ultimate death [3]. Untimely malarial diagnosis unnecessarily complicates the line of treatment and causes additional adversities including drug resistance, chemopreventive abuse, and genetic mutations which are physically, mentally, and monetarily catastrophic to handle [4]. Therefore, timely, and accurate diagnosis of malaria is of critical importance.

Giemsa is a purple-colored stain that explicitly highlights the parasite in thin slides of the patient's blood [5]. Parasitologists traditionally examine the stained blood cell smears for the purple-stained parasite to confirm malarial infiltration. However, the degree of accuracy of diagnosis is largely variable and subject to the expertise of the parasitologist, the contrast, illumination and brightness of smear images obtained, as well as the quality of the testing equipment used. This may lead to discrepencies in the the diagnosis pipeline, ultimately leading to misdiagnosis [6]. To effectively shorten the detection pipeline while improving the accuracy of the diagnosis, the computer-aided diagnostic landscape was introduced. This automated the detection process and improved the speed, accuracy and cost-effectivity of the line of treatment. Several methods were introduced to develop efficient healthcare delivery systems for computer-aided malarial diagnosis.

Image Processing techniques were introduced to remove pertaining noise, correct illumination, normalize edge segmentation, and eradicate artifacts from cell images to render them useful for automated examination. Mean [7], Median [8], Geometric Mean [9], and Wiener filtering [10] are notable image denoising methods that operate in the pixel's neighbourhood for image impulse noise removal and edge preservation. Savkare et al. [11] introduced a novel method to enhance malarial image resolution through Laplacian filtering and Adaptive Histogram Equalization. Razzak et al. [12] proposed top-hat transforms for correcting irregularity in image illumination. Dong et al. [13] suggested morphological operations for artifact removal and hole filling for substituting the missing frequencies in a signal with appropriate values. Once the images have been preprocessed, a segmentation algorithm is leveraged for highlighting parasitic infestation. Threshold-based segmentation algorithms such as Otsu's thresholding were amalgamated with morphological operations such as granulometry to calculate optimum threshold values for bimodal images. Watershed Segmentation for Malaria was introduced by Kim et al. [14] for extracting the boundaries of the parasite inside the cell. Marker-controlled Watershed [15] is an improvement over the traditional Watershed segmentation as it includes the segmentation of extrinsically overlapping cells as well. Bibin et al. [16] introduced Active Contour Models for malarial cell segmentation by topological levelling. Abbas et al.'s Gaussian Mixture Model [17] for malaria identification worked with the probabilistic assumption that all the defined data points arose from a mixture of Gaussian distributions. Image Processing models are simple to implement, but are only suitable for smaller-sized datasets. Learning-based methods return better results for larger-sized datasets.

Learning-based methods could be either Supervised or Unsupervised. Unsupervised learning methods such as the Naïve Bayes Tree by Das et al. [18] return substantially satisfactory results without providing training output data. However, Supervised learning methods such as the Random Forest Classifier proposed by Quinn et al. [19] return better classification accuracy. Neural Networks have proven to be remarkably efficient for modeling image data. Gopakumar et al. [20] proposed a Convolutional Neural Network (CNN) to operate on the focal parasitemia of blood smear images. However, training CNNs straight from scratch is a computationally intensive and time-consuming process.

In this paper, we propose ResNet50, a deep neural network (DNN) for classifying malarial cell images into the parasitized and uninfected classes. Since building a DNN from the ground-up is expensive, we introduce Transfer Learning (TL). TL fine-tunes pre-trained models to accommodate alien weights without having to train the model from absolute scratch.

**Materials and Method**

The open-access dataset provided by the National Library of Medicine's Lister Hill National Center for Biomedical Communication [21] has been used in this paper. It was created in collaboration with the National Institutes of Health, the Centers for Disease Control and Prevention, and the Mahidol-Oxford Tropical Medicine Research Unit, and is now considered a benchmark dataset for computer vision tasks aimed at classifying malarial cell images.

The dataset contains 27,558 thin blood smear slide images clicked using light microscopy of 150 P *falciparum* malaria-infected and 50 healthy individuals. The dataset is equally distributed into the parasitized and uninfected classes, each containing 13,779 images, with a training, testing and validation ratio of 7:2:1.

Before implementing the classification algorithm, data augmentation is done to make the dataset more robust to image quality variations. MixUp [22] and CutMix [23] are the two algorithms used for data augmentation. MixUp involves randomly combining pairs of images and labels by taking a weighted linear combination of the two, while CutMix involves cutting and pasting random patches from one image to another with corresponding label adjustments. Both techniques aim to improve model generalization by encouraging smoother and more robust decision boundaries. Fig.1 illustrates the output images of MixUp and CutMix augmentation. After the dataset has been augmented, it is passed through the ResNet50 classifier.

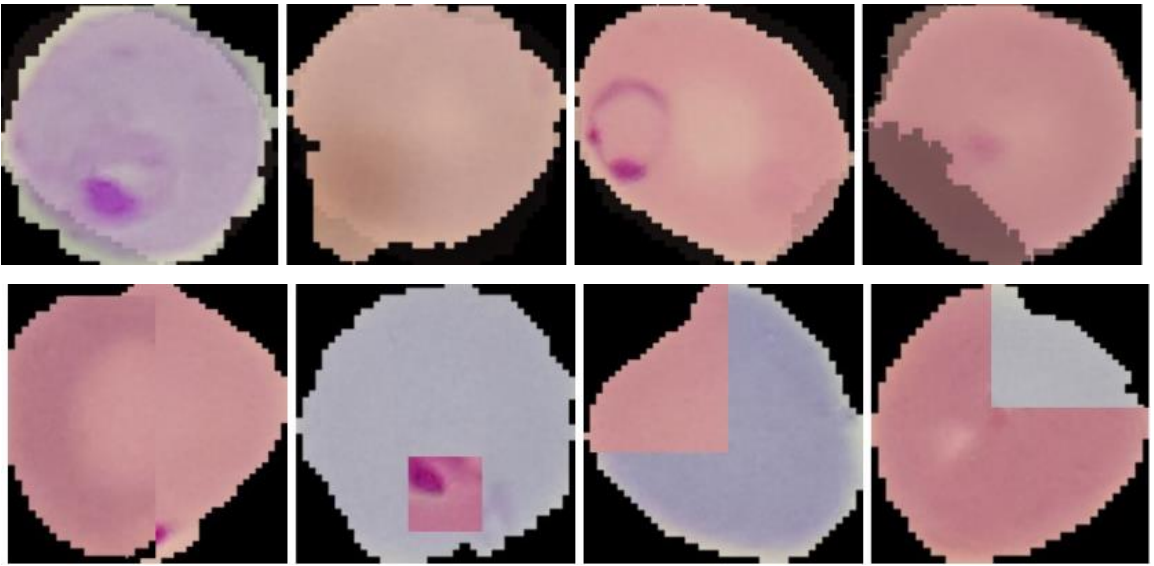

**Fig 1. Results of MixUp and CutMix Augmentation**

ResNet50 [24] is a member of the deep neural network family that is known as "Residual Networks". These networks have special connections called "residual connections" that permit the network to bypass certain layers and "skip connections" that enable gradients to flow more smoothly through the network. ResNets are highly effective in binary image classification tasks, especially when using the sigmoid activation function which is known to experience the vanishing gradient problem. ResNets tackle this issue by using the residual connections to create shortcuts from the input to the output, which bypasses layers and allows gradients to flow more effectively during backpropagation. ResNet50 contains 49 convolutional layers followed by a fully connected layer at the end. Fig 2. Illustrates the architecture of the ResNet50 backbone. The convolutional layers are organized into four blocks, each with a different number of layers and downsampling operations. The first block has three layers and downsamples the spatial dimensions of the input cell image by a factor of 4, while the other three blocks each have four layers and downsample the input image by a factor of 2. The final fully connected layer maps the output of the last convolutional layer to the desired output dimensionality. The model was trained for 150 epochs on the NVIDIA Tesla P100 GPU. The Adam optimizer was used with the Binary cross entropy loss function for the training. The Sigmoid activation function is used in the classification layer.

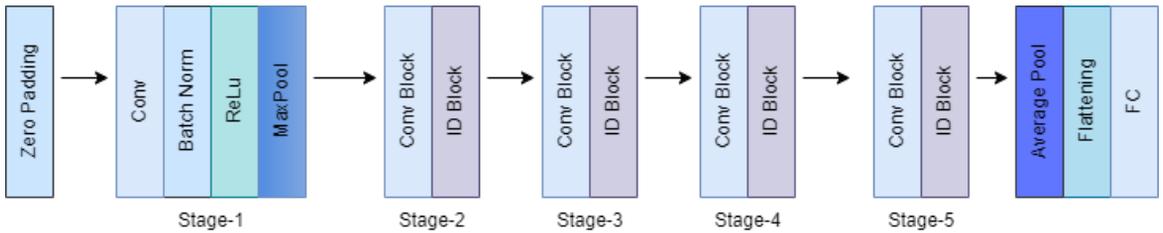

**Fig 2. Architecture of ResNet50**

**Results**

The performance of the ResNet50 model is compared with 3 other state-of-the-art models that have been used to detect malaria in cell images. The models chosen for comparison belong to three different classes of classifiers used in medical imaging. Watershed segmentation is an example of an image processing-based thresholding technique for identifying boundaries of the parasite. Random Forest is a tree-based machine learning algorithm which culminates the outputs of multiple decision trees to reach the result. VGG16 [25] belongs to the VGG group of deep neural network models trained on the ImageNet dataset for image classification. It has 16 trainable layers.

The models have been juxtaposed based on 3 performance metrics. These metrics; accuracy, precision and recall have been used to compare the performance of the model.

Accuracy is used to show the number of correctly classified malarial cells over the total number of cells. It can be represented as:

$$Accuracy\ (\%) = \frac{TP + TN}{TP + TN + FP + FN} \times 100 \quad (1)$$

Precision is determined by dividing the genuine positives by any positive prediction. It measures how well a model can predict a particular category.

$$Precision = \frac{TP}{(TP+FP)} \quad (2)$$

Recall is the true positive rate of any model. It is computed by dividing the real positives by anything that has been predicted as positive, rightly or not.

$$Recall = \frac{TP}{(TP+FN)} \quad (3)$$

Table 1. Performance Metrics

| Model Name | Accuracy | Precision | Recall |
|---|---|---|---|
| Random Forest | 0.651 | 0.740 | 0.740 |
| VGG16 | 0.937 | 0.529 | 0.744 |
| Watershed Segmentation | 0.90 | 0.643 | 0.662 |
| **Proposed Model** | **0.9875** | **0.993** | **0.995** |

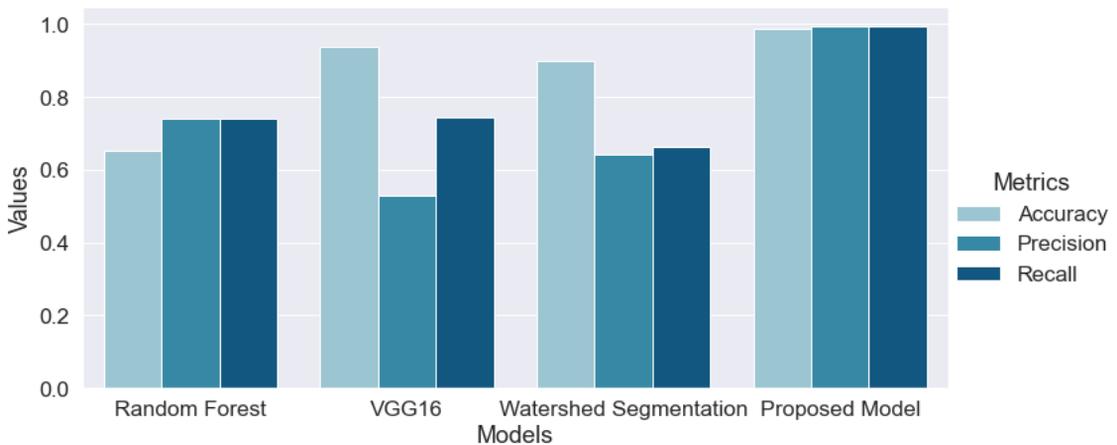

Fig 3. Performance Metrics for the four models

Discussion

It can be observed from Table 1 and Fig 3 that the proposed model returned the best accuracy, precision, and recall values of 98.75%, 99.3% and 99.5%. VGG16 returned comparable accuracy of 93.7%, which is acceptable but not as decent as the proposed model's. It has lower precision and recall values of 52.9% and 74.4%, indicating that the model might have overfitted. This could be a result of the vanishing gradient problem associated with the Sigmoid activation function. Random Forest returned 65.1% accuracy, with 74% precision and recall values. These are extremely underperforming metrics, especially when compared to the proposed model. Watershed segmentation returned 90% segmentation accuracy, with 64.3% precision and 66.2% recall values. The poor precision and recall values are a primary by-product of the dataset's large size. Fig.4. illustrates the true labels and the predicted labels for a set of malarial cells. It can be seen how the ResNet50 backbone returned accurate labels for each of the test samples. Hence, it can be observed that the proposed model outperforms all the 3 models used for comparison.

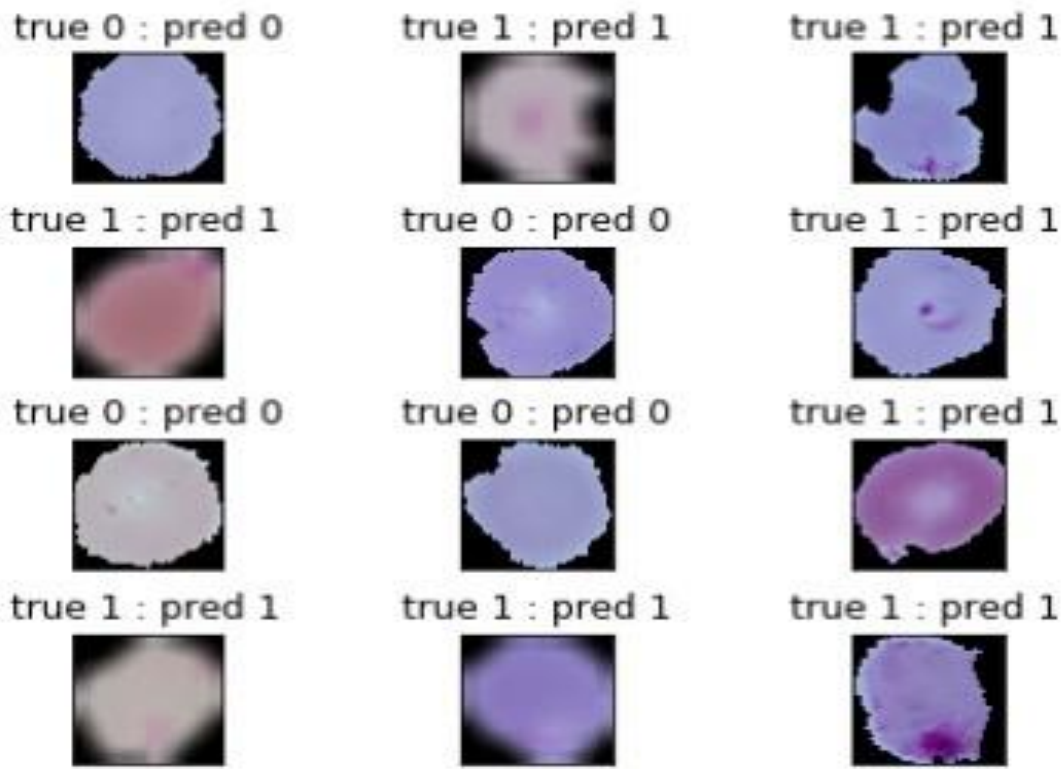

Fig 4. Classification Output of ResNet50 model

**Conclusion**

Therefore, it is concluded that malaria detection could be faster and more accurate due to TL-based methods. ResNet50 showed the best binary classification results when compared with other thresholding-based, learning-based, and neural network-based technqiues. The only problem associated with ResNets is that a large amount of training data is required to achieve optimal performance. While the malaria dataset had an adequate number of images per class, this method could not be expanded to other diseases with smaller-sized datasets. In our future work, we aim to leverage Vision Transformers for enabling disease detection on medical datasets with a lesser number of images.